\documentclass[
prb,twocolumn,aps]{revtex4-1}

\usepackage{graphicx}
\usepackage{dcolumn}
\usepackage{bm}
\usepackage{color}

\newcommand{\bea}{\begin{eqnarray}}
\newcommand{\eea}{\end{eqnarray}}

\newcommand{\bk}{{\bf k}}

\DeclareMathAlphabet{\mathpzc}{OT1}{pzc}{m}{it} 

\def\be{\begin{equation}}
\def\ee{\end{equation}}
\def\ba{\begin{eqnarray}}
\def\ea{\end{eqnarray}}

\def\YBCO{YBa$_2$Cu$_3$O$_{7-\delta}$}

\def\C60{A$_x$C$_{60}$}

\begin{document}
\title{
Fermi Pockets in a d-wave Superconductor with Coexisting Loop-Current Order }
\author{ S. A. Kivelson}
 \affiliation{Department of Physics, Stanford University, Stanford, CA 94305}
 \author{ C.M. Varma}
\affiliation{Department of Physics, University of California Riverside, Riverside, CA 92521}
\date{\today}
\begin{abstract}
A d-wave superconducting phase with coexisting intra-unit-cell orbital current order has the remarkable property that it supports finite size Fermi pockets of Bougoliubov
 quasiparticles.  Experimentally detectable consequences of this include a residual $T$-linear term in the specific heat {\it in the absence of disorder} and residual features in the thermal and microwave conductivity in the low disorder limit. 
\end{abstract}
\maketitle

As a theoretical proposition, it has recently been realized theoretically\cite{BergKivelson,Orgad,Kallin,spinliquid} that there exist superconducting phases in which there is a stable residual Fermi surface and correspondingly a finite density of states in the limit $E\to 0$.  One case in which this occurs, which will be discussed explicitly in this paper, is the case of coexisting d-wave superconductivity and intra-unit cell orbital loop current order of the type proposed\cite{Varma,simon}, 
 as the underlying cause of the pseudo-gap regime in the cuprate high temperature superconductors.  Other interesting examples are the ``pair-density wave''\cite{Bergpdw} and colinear FFLO states,\cite{Radzihovsky2} which involve condensation of Cooper pairs with non-zero center of mass momentum.\cite{Orgad,Kallin,Radzihovsky}

Two 
experimental developments are suggestive that this sort of 
behavior may be relevant in real materials:  

\noindent{1) }  Riggs {\it et al} \cite{Riggs} have noted the existence of a residual $T$-linear term in the specific heat in high quality crystals of the cuprate superconductor, {\YBCO}, with a magnitude $C/T \approx 2$mJ/mole-K$^2$ that appears to approach a non-zero value in the limit of vanishing disorder. Related to this is the fact that there is a finite paramagnetic conductivity at low frequencies in the superconducting state in the limit $T \to 0$ for samples with very low disorder \cite{Harris}.   If these are borne out in future experiments, it is worth exploring the possibility that it is due to a residual Fermi surface {\it in the superconducting state}.  
  
  \noindent{2)}  A set of anomalies\cite{dmit,dmit2} have been reported in certain insulating crystals with nearly triangular lattice structures and an odd integer number of electrons per unit cell.  These materials show no evidence of magnetic order down to temperatures well below the scale of the exchange interactions, and although they are insulating, they exhibit several 
  Fermi-liquid-like low temperature behaviors including a $T$ linear specific heat, a $T$ independent Pauli susceptibility, and a $T$-linear thermal conductivity.   It is an exciting possibility that this behavior reflects the existence of a spin-liquid phase, but there is no consensus on exactly what type of spin liquid.  One possible set of spin-liquids which have been proposed in this context are formally analogous to superconducting phases, with ``spinon pairing'' playing the role of electron pairing.  
 Two particular such proposals  which have been considered in this context are the spin-liquid analogue of a pair-density-wave state\cite{leelee} and a state with d-wave spinon pairing coexisting with a form of loop current order\cite{spinliquid}, analogous to the state considered in the present paper.

\section{Fermi pockets in a current loop ordered superconductor}
Typically in a superconducting state, the fact that the superconducting gap perfectly nests the Fermi surface suggests the existence of a gap on the entire Fermi surface, or at most the existence of gapless nodal points (in 2D) or nodal points or nodal lines in 3D.   
However, several states are now known which admit a residual Fermi surface in the superconducting state. 

\begin{figure}
\centerline{\includegraphics[clip,width=3.0in]{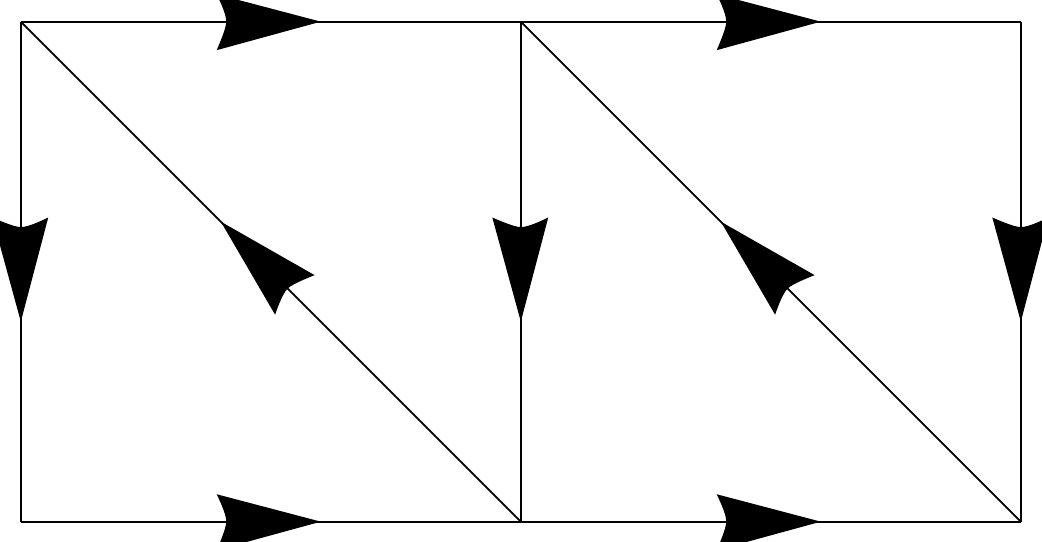}}
\caption{
Schematic representation of the pattern of ground-state currents in an orbital current ordered state.}
\label{currents}
\end{figure}

To be explicit, with the cuprates in mind, we consider 
 a simple single orbital model on a square lattice which mimics the symmetry of the loop-current order derived in the mean-field approximation from the three-orbitals per unit-cell model. The non-interacting band dispersion is $\epsilon_{\bf k}$ and there is a superconducting gap with d-wave symmetry, $\Delta_{\bf k} = \eta_d({\bf k})|\Delta_{\bf k}|$ (where $\eta_d(\vec k) = {\rm sign}(k_x^2-k_y^2)$ is the d-wave form-factor). Both $\epsilon_{\bf k}$ and $|\Delta_{\bf k}|$ respect all the lattice symmetries, time reversal symmetry, and (for simplicity) spin rotational symmetry.  
 The band dispersion is altered, $\epsilon_{\bf k}\to \epsilon_{\bf k}+J_{\bf k}$, in the loop current state by a mean-field shift in the dispersion  with magnitude proportional to the symmetry breaking order parameter.  The orbital loop current order we are considering (shown schematically in Fig. \ref{currents}) preserves spin rotational symmetry and is even under $(x,y) \to (-y,-x)$, but is odd under time-reversal and under the reflection $(x,y) \to (y,x)$.  From this it follows that $J_{k_x,k_y} =-J_{-k_x,-k_y}= -J_{k_y,k_x}$ and thus that the quasi-particle dispersion is
\bea
E_{\bk}=J_\bk \pm \sqrt{\epsilon_\bk^2+|\Delta_\bk|^2}.
\eea

\begin{figure}
\centerline{\includegraphics[clip,width=2.5in]{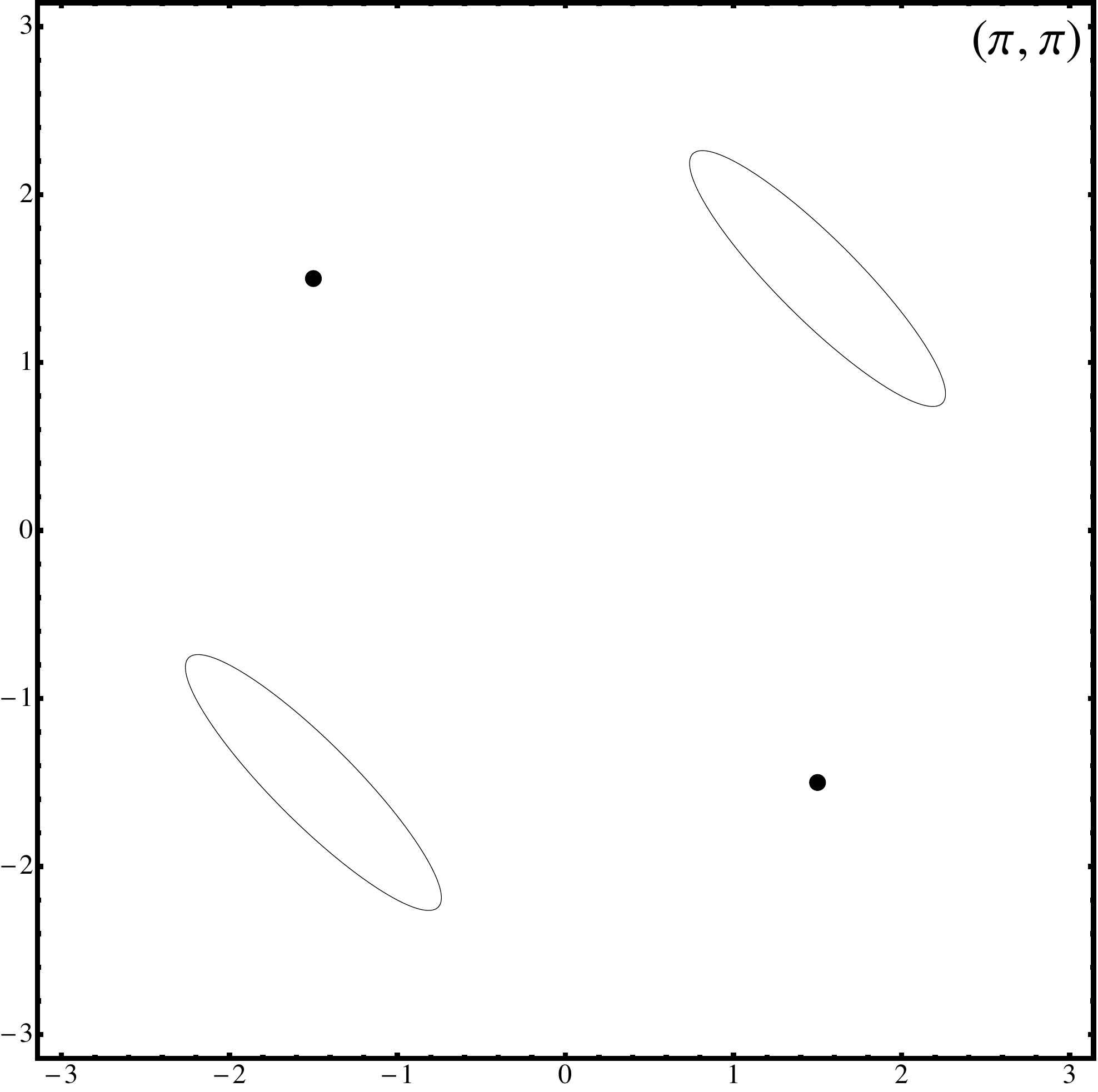}}
\caption{
Fermi surface in the first Brillouin zone of a d-wave superconductor with coexisting intra-unit cell loop order of the sort discussed in the text.  The Fermi surface consists of two elliptical pockets along the $(1,1)$ direction and two nodal points along the $(1,-1)$ direction.
}
\label{fig-FS.pdf}
\end{figure}

As long as the Fermi surface crosses the zone diagonal, $k_x =\pm k_y$, the pure d-wave superconducting state possesses gapless nodal points at $k_x = k_y= \pm q_{node}$ and $k_x = -k_y= \pm q_{node}$. It is important to note that reflection symmetry implies $J(q,q)=0$, but, in general, $J(q,-q)$ is non-zero, and in particular that $J(q_{node},-q_{node})= -J(-q_{node},q_{node})\equiv \bar J$ where $\bar J$ is how we will parameterize the magnitude of the current loop order.  Thus, as shown in Fig. \ref{fig-FS.pdf}, the nodes along the $(1,1)$ direction are entirely unaffected by weak loop current order, while the nodes along the $(1,-1)$ direction are shifted away from the Fermi energy.  In particular, for $\bk=\Lambda [q_{node} + \delta k_\perp](1,-1) + \delta k_\parallel (1,1)$ with $\Lambda=\pm 1$, 
\bea
E_\bk = \Lambda \bar J \pm \sqrt{(v_F\delta k_\perp)^2 +(v_\Delta \delta k_\parallel)^2} +\ldots
\eea
where $v_F$ and $v_\Delta$ are the appropriate derivatives of $\epsilon_\bk$ and $\Delta_\bk$, respectively, and $\ldots$ signifies higher order terms in powers of $\delta k$.  Note that one node is shifted upward relative to the chemical potential and the other is shifted down by the same amount.  The result is the Fermi surface shown in Fig. \ref{fig-FS.pdf} with two nodal points and two 
elliptical Fermi pockets.  The resulting density of states to linear order in the energy $E$ (measured relative to the Fermi energy) is easily seen to be
\bea
N(0) = \frac 2 \pi \frac {\left(\bar J + |E|\right)} {\hbar^2v_Fv_\Delta}.
\label{DOS}
\eea
where the first term comes from the Fermi pockets and the second 
 from the two remaining nodal points.

It is important to stress that the quasiparticles in question are not the quasiparticles of a Fermi liquid, but rather the Bogoliubov quasiparticles of a superconductor.  For instance, one can readily see that the expectation value of the charge of a quasiparticle with crystal momentum ${\bf k}$ is 
\bea
Q_\bk =e \frac {\epsilon_\bk}{ \sqrt{\epsilon_\bk^2+|\Delta_\bk|^2}} \ \to \ e\frac {\epsilon_\bk}{|J_\bk|} 
\eea
where the second expression is valid for states of zero energy ({\it i.e.} on the Fermi pocket).  Thus, it is easily seen that the quasiparticle charge is electron-like for $\bk$ outside the underlying Fermi surface, hole-like inside the Fermi surface, and vanishes wherever the Fermi pocket crosses the underlying Fermi surface.   For small magnitude of $\bar J$, where the spectrum can be linearized in the neighborhood of the nodal point, the average of $Q_\bk$ along the Fermi pocket vanishes, but the mean squared charge is
\bea
\left[Q^2\right] = e^2\left[1 -{\cal O}(\bar J/\Delta) \right].
\label{averagecharge}
\eea

 Note that {\it at least} certain quantitative aspects of these results may be different in more complex models of orbital current ordered states.  For example, it has been argued\cite{he} that quantum fluctuations between the four symmetry related classical states (of which the example shown in Fig. 1 is one) play an unusual role in the   physics of the state.  Moreover, aspects of the electronic structure associated with multiple atoms per unit cell could lead to new features of the ordered state which cannot be captured in the context of  single-band models. 

As a point of comparison, for spin polarized electrons in a ferromagnet, the analysis is very similar.  Here the quasiparticle dispersion becomes
\bea
E_{\bk}=h\sigma \pm \sqrt{\epsilon_\bk^2+|\Delta_\bk|^2}.
\eea
where $h$ is the effective Zeeman field and $\sigma=\pm 1$ is the spin-polarization of the quasiparticle.  Now  the nodal points for spin-up electrons are shifted down below the chemical potential, and those for spin down electrons are shifted up, so that there are four Fermi pockets.

\section{Measurable quantities}
Here we obtain theoretical expressions for some of the most anomalous experimentally measurable consequences of the existence of a Fermi pocket in a superconducting state with coexisting d-wave superconducting and orbital current order.  We compare the results to those expected for a d-wave superconductor without coexisting loop current order, both with or without the effects of weak impurity scattering taken into account.

From the expression in Eq. \ref{DOS}, it follows  that the quasiparticle contribution to the specific heat at low $T$ is
\bea
\frac C T\equiv \gamma_{loop} = \frac {2\pi k_B^2} {3\hbar^2v_Fv_\Delta} \bar J \left [1+ 27\zeta(3)\left( \frac {k_BT} {\bar J} \right)+\ldots\right],
\label{gammaloop}
\eea
where the first (leading order) term comes from the pockets and the second (subleading) term from the point nodes and $\ldots$ signifies higher order terms in powers of $k_BT/|\bar J|$.  (Here $\zeta(3) =1.202 \ldots$ is the Reiman Zeta function.)  In  the absence of orbital current order and in the absence of disorder, $\gamma$ vanishes as $T\to 0$.  However\cite{schmitt,peter}, in the presence of disorder (at least to the extent that scattering can be treated in t-matrix approximation), the nodal density of states is broadened on the scale of the elastic scattering rate, $\hbar/\tau$ 
and consequently for $T \to 0$,
\bea
\gamma_{nodal} \approx  \frac {(\pi^2/3)k_B} {v_Fv_\Delta} \frac \hbar \tau \ln\left|\frac{\Delta\tau}{\hbar}\right|
\eea
In comparing the effect of disorder with that  of loop current order, note that $\gamma_{loop}$ is not only independent of $\tau$ in the weak disorder limit, it is parametrically larger than $\gamma_{nodal}$ by a factor $O(\bar J\tau/\hbar)$.

A well known signature of a nodal superconductor, at least in the strongly Type II limit (in which $H_{c1}$ is negligibly small), is the singular magnetic-field dependence of the low temperature specific heat\cite{Volovik,V2}.  
 While even in a fully gapped superconductor there is a non-analytic contribution to the specific heat proportional to $|H|$, {\it i.e.} to the density of vortices, there is an even stronger field dependence in a nodal superconductor due to the doplar shift of the quasiparticle spectrum produced by the superfluid flow at long distances from the vortex core.  Consequently, in a pure d-wave superconductor,
\bea
\gamma_{nodal}(H) - \gamma_{nodal}(0) \sim  \frac { k_B^2}  {\hbar v_\Delta}\sqrt{\frac {|H|}{\phi_0}} + \ldots
\label{gammanodal}
\eea
where $\phi_0=hc/e$.
However, in the presence of scattering, this strong field dependence is suppressed\cite{peter} for $|H| < \phi_0/(v_F \tau)^2$ . Thus, if a residual $\gamma$ is produced by impurity scattering, a corresponding rounding of the $\sqrt{H}$ field dependence of the specific heat is expected as well.

In a d-wave superconductor with coexisting loop current order, the leading order contribution to $\gamma$ is the $|H|^0$ term, already computed in Eq. \ref{gammaloop}, which comes from the pockets.  However, the next to leading term comes from the remaining nodal points, and is thus of precisely of the same form as in Eq. \ref{gammanodal}, but with half the magnitude.  A substantial residual value of $\gamma$ as both $H$ and $T$ tend  to zero, coexisting with a $\sqrt{H}$ contribution to $\gamma$ which persists to asymptotically low fields would constitute strong evidence of loop current order (or a  
relative).

The non-vanishing density of states in the loop ordered state has  consequences for transport properties as well.  In general, we expect the thermal conductivity to obey the relation
\bea
\kappa_{nodal} = (1/2) C(T) <v^2>\tau =T\gamma<v^2>\tau,
\eea 
where $<v^2>$ is the mean-squared quasiparticle velocity.
For the pure d-wave superconductor with weak scattering, the factor of $1/\tau$ in $\gamma$ cancels the explicit factor of $\tau$ in the expression for $\kappa$ producing a ``universal'' expression\cite{fradkin,durstandlee} for $\kappa$, {\it i.e.} one that is independent of $\tau$.  The parametrically larger magnitude of $\gamma$ in the loop ordered state results in a parametrically larger magnitude of $\kappa$ in the clean limit of the loop ordered state
\bea
\frac {\kappa_{loop}} {\kappa_{nodal}} \sim\frac { |\bar J|\tau_{loop}}{\hbar}
\eea
Note that
$\tau_{loop}$ must be calculated for the loop current state and not for the nodal points. 

The conductivity may be estimated by  an approximate version of the Wiedemann-Franz law  
\bea
\sigma(0)=\frac {3e^2}{(\pi k_B)^2}\left( \frac \kappa T\right)
\eea
apart from possible different Fermi-liquid renormalizations and vertex corrections. Extension of this at $T\to 0$ but finite frequencies should lead to  an expression for the microwave  conductivity 
\bea
\sigma(\omega) \approx \frac{ \sigma(0)}{1 + \omega^2 \tau_{loop}^2}.
\eea 
The integral over $\omega$ of this should then subtract from the superfluid density or $\lambda_L^{-2}$, where $\lambda_L$ is the London penetration depth. 

An obvious issue is whether the existence of Fermi pockets in the superconducting state produces anything resembling the periodic oscillations as a function of $1/H$, as would be expected in a Fermi liquid with similar dispersion, {\it i.e.} with a frequency, $H^*=( hc/2\pi e) A$, where $A$ is the area of the Fermi pocket in the BZ.  On the other hand, as discussed above, the charge changes sign on different portions of the Fermi surface such that the average charge of the quasiparticles is close to 0. 
However, Andreev scattering at the edges of the pockets where the charge is $0$ mixes the particle and hole states such that closed orbits in a magnetic field are in principle possible. 
Evidence that such oscillations occur has been reported by Kallin {\it et al}\cite{Kallin} in the case in which the Fermi pockets arise in a pair-density wave.  For pockets arising from orbital loop current order, Wang and Vafek\cite{wangvafek} have found oscillations of the specific heat with a frequency roughly proportional to $A$, although the field and temperature dependence appears rather more complex than for a Fermi-liquid.  On the otherhand, Senthil {\it et al}\cite{Senthil}, in a different formulation of the problem of magneto-oscillations in the superconducting state do not find such oscillations.  Analytic understanding of these results is presently lacking, although this is clearly an important and interesting problem.

\section{Discussion}

The existence of a well defined Bogoliubov Fermi surface in certain exotic superconducting states is a relatively new theoretical development.  What is clear is that it likewise gives rise to complex emergent behavior at low temperatures, including characteristic  variations of the specific heat (and, presumably, other measurable quantities) which reflect the low energy electronic structure.

We conclude with a few comments  concerning the possible relation of our present results with experiments in the cuprate superconductor, YBCO.  As mentioned in the introduction, one puzzling feature of the specific heat in YBCO is that there is a non-zero value of $\gamma(0,0)\approx 2 mJ/mole K^2$ in the superconducting state of underdoped materials which does not appear to depend greatly on sample quality, and so may be intrinsic.  This is compatible with the existence of a Fermi surface of BdG quasiparticles.  If this is interpreted as being the result of coexisting loop order, then taking  values of $v_F \approx 2 \times 10^5 m/s$ and $v_\Delta/v_F \approx 1/8$ deduced in experiments, one can work backward from this to obtain a value for $\bar J \approx 60meV$ from the expression in Eq. \ref{gammaloop}.

 We note that the anomalous paramagnetic conductivity \cite{Harris} observed in the superconducting state may also be 
 compatible with this explanation. In the superconducting state an integrated paramangetic conductivity $S$ of about $2\pm 1/2 \% $ of the normal state value is observed in YBa$_2$Cu$_3$O$_{6.5}$, which decreases to $1/2 \pm 1/2\%$ for YBa$_2$Cu$_3$O$_{6.99}$.  $S$ is proportional to the number of electrons in the pockets, {\it i.e.} $S \sim (1/2\pi) (a^2\bar J^2/\hbar^2v_Fv_\Delta)$. With the $\bar J$ inferred from the specific heat, we would estimate it to be about $3\%$ and would expect it to go to $0$ as loop order disappears in the phase diagram. 
The significance of the microwave conductivity experiments is also that the finite temperature changes of the integrated weight parallel the measured change of the superfluid density with temperature, which is linear. This puts additional weight to the argument that the residual microwave conductivity and specific heat are not extrinsic effects due to impurities. 
Assuming that surviving Fermi pockets are, indeed, the origin of the residual specific heat and the uncondensed spectral weight\cite{Harris}, one infers that $\tau_{loop} \approx  10^{-11}$ secs. 
This, in turn, would imply a huge enhancement of the thermal conductivity over the ``universal value"  
 for similar tempratures and samples.  This does not appear to have been seen, so far\cite{taillefer}.
 
 There are many further experimental tests that  can be implemented to test for the existence of near-nodal pockets in the superconducting state. In terms of bulk measurements, the most promising are systematic measurements of the field independent specific heat and paramagnetic conductivity at low temperature in the superconducting state as a function of doping.
 Turning to surface probes, high resolution ARPES measurements  of the Fermi-surface and STM studies of quasiparticle interference patterns  could, potentially provide the most direct evidence for or against the existence of nodal pockets {\em in the superconducting state}. 
 Note that this is an entirely separate issue from the issue from the origin of the ``Fermi arcs'' in the ``pseudo-gap'' state above $T_c$.

 \begin{acknowledgements}
SK acknowledges extremely important conversations with Oskar Vafek and Luyang Wang.
CMV wishes to thank the Physics and Applied Physics faculty  for their warm reception and the Stanford Institute for Theoretical Physics for financial support. 
We thank S. Lederer for providing Fig. 2.
 This work was supported in part by NSF Grant No. DMR-0906530 (CMV) and DOE grant No. DE-AC02-76SF00515 (SAK).
\end{acknowledgements}
\bibliography{loop}
\end{document}